\documentclass[twocolumn,amssymb,showpacs]{revtex4}

\usepackage{graphicx}
\usepackage{color}
\usepackage{epsfig}
\usepackage{latexsym}
\usepackage{bm}
\usepackage{ulem}

\begin{document}

\renewcommand{\ni}{{\noindent}}
\newcommand{\dprime}{{\prime\prime}}
\newcommand{\be}{\begin{equation}}
\newcommand{\ee}{\end{equation}}
\newcommand{\bea}{\begin{eqnarray}}
\newcommand{\eea}{\end{eqnarray}}
\newcommand{\nn}{\nonumber}
\newcommand{\bk}{{\bf k}}
\newcommand{\bQ}{{\bf Q}}
\newcommand{\q}{{\bf q}}
\newcommand{\s}{{\bf s}}
\newcommand{\bN}{{\bf \nabla}}
\newcommand{\bA}{{\bf A}}
\newcommand{\bE}{{\bf E}}
\newcommand{\bj}{{\bf j}}
\newcommand{\bJ}{{\bf J}}
\newcommand{\bs}{{\bf v}_s}
\newcommand{\bn}{{\bf v}_n}
\newcommand{\bv}{{\bf v}}
\newcommand{\la}{\langle}
\newcommand{\ra}{\rangle}
\newcommand{\dg}{\dagger}
\newcommand{\br}{{\bf{r}}}
\newcommand{\brp}{{\bf{r}^\prime}}
\newcommand{\bq}{{\bf{q}}}
\newcommand{\hx}{\hat{\bf x}}
\newcommand{\hy}{\hat{\bf y}}
\newcommand{\bS}{{\bf S}}
\newcommand{\cU}{{\cal U}}
\newcommand{\cD}{{\cal D}}
\newcommand{\bR}{{\bf R}}
\newcommand{\pll}{\parallel}
\newcommand{\sumr}{\sum_{\vr}}
\newcommand{\cP}{{\cal P}}
\newcommand{\cQ}{{\cal Q}}
\newcommand{\cS}{{\cal S}}
\newcommand{\ua}{\uparrow}
\newcommand{\da}{\downarrow}
\newcommand{\red}{\textcolor {red}}
\newcommand{\blu}{\textcolor {blue}}
\newcommand{\1}{{\oldstylenums{1}}}
\newcommand{\2}{{\oldstylenums{2}}}
\newcommand{\mDelta}{\varepsilon}
\newcommand{\m}{\tilde m}
\def\lsim {\protect \raisebox{-0.75ex}[-1.5ex]{$\;\stackrel{<}{\sim}\;$}}
\def\gsim {\protect \raisebox{-0.75ex}[-1.5ex]{$\;\stackrel{>}{\sim}\;$}}
\def\lsimeq {\protect \raisebox{-0.75ex}[-1.5ex]{$\;\stackrel{<}{\simeq}\;$}}
\def\gsimeq {\protect \raisebox{-0.75ex}[-1.5ex]{$\;\stackrel{>}{\simeq}\;$}}

\

\title{ Zeroth law and nonequilibrium thermodynamics for steady states in contact}

\author{Sayani Chatterjee$^{1}$, Punyabrata Pradhan$^{1}$ and P. K. 
Mohanty$^{2,3}$}

\affiliation{ $^1$Department of Theoretical Sciences, S. N. Bose 
National Centre for Basic Sciences, Kolkata 700098, India \\ 
$^2$CMP Division, Saha Institute of Nuclear Physics, 1/AF Bidhan 
Nagar, Kolkata 700064, India \\ $^3$Max Planck Institute for the 
Physics of Complex Systems, 01187 Dresden, Germany}

\begin{abstract}
\noindent{ We ask what happens when two nonequilibrium systems 
in steady state are kept in contact and allowed to
exchange a quantity, say mass, which is conserved in the combined
system. Will the systems eventually evolve to a new stationary
state where certain intensive thermodynamic variable, like
equilibrium chemical potential, equalizes following zeroth law
of thermodynamics and, if so, under what conditions is it possible?
We argue that an equilibrium-like thermodynamic structure can be 
extended to nonequilibrium steady states having 
short-ranged spatial correlations, provided that the systems {\it 
interact weakly} to exchange mass with rates satisfying a balance 
condition - reminiscent of detailed balance condition in 
equilibrium. The short-ranged correlations would lead to subsystem 
factorization on a coarse-grained level and the balance condition 
ensures both equalization of an intensive thermodynamic variable as 
well as ensemble equivalence, which are crucial for construction of a well-defined nonequilibrium thermodynamics. This proposition is 
proved and demonstrated in various conserved-mass transport 
processes having nonzero spatial correlations.
}

\typeout{polish abstract}
\end{abstract}


\maketitle

\section{Introduction}

Zeroth law is the cornerstone of equilibrium
thermodynamics. It states that, if two systems are separately in
equilibrium with a third one, they are also in equilibrium with
each other \cite{Kardar}. An immediate consequence of the zeroth
law is the existence of state functions - a set of intensive
thermodynamic variables (ITV) which equalize for two systems in
contact. For example, if two systems are
allowed to exchange a conserved quantity, say mass, they
eventually achieve equilibrium where chemical potential becomes
uniform throughout  the {\it combined} systems. The striking
feature of this thermodynamic structure is that all equilibrium
systems form equivalence classes where each class is specified by
a particular ITV. Then a system, an element of a particular class,
is related to any other system in the class by a property that
they have the same value of the ITV.

We ask whether a similar thermodynamic characterization is
possible in general for systems having a nonequilibrium steady 
state (NESS). Can equalization of an ITV, governing 
``equilibration'' between two steady-state systems in contact, be 
used to construct such equivalence classes? The answer is 
nontrivial; in fact, it is not even clear if such a formulation is 
at all possible \cite{Casas, Eyink1996, Kurchan_PRE1997, 
OonoPaniconi, Barrat_PRL2000, Baranyai_PRE2000, 
Behringer_Nature2002, HayashiSasa_PRE2003, 
SasaTasaki_JStatPhys2006, Sasa2014, Dickman_PRE2014}. In this
paper, we find an affirmative answer to this question, which can
lead to a remarkable thermodynamic structure where a vast
class of systems having a NESS form equivalence classes, 
equilibrium systems of course included.

There have been extensive studies in the past to find a suitable
statistical mechanical framework for systems having a NESS
\cite{Casas, Eyink1996, OonoPaniconi, Bertini_PRL2001,
HayashiSasa_PRE2003, SasaTasaki_JStatPhys2006, Sasa2014,
Dickman_PRE2014, Sasa_PRL2008, Komatsu_PRL2008, Bertin_PRL2006}.
Though the studies have not yet converged to a universal picture,
it has been realized that suitably chosen mass exchange rates at
the contact could possibly lead to proper formulation of a nonequilibrium thermodynamics \cite{SasaTasaki_JStatPhys2006,
Sasa2014, Bertin_PRL2006, Bertin_PRE2007, Pradhan_PRL2010,
Pradhan1_PRE2011}. An appropriate contact dynamics is crucial
because, without it, properties of mass fluctuations in a system
would be different, depending on whether the system is in contact
(grandcanonical) or {\it not} in contact (canonical) with other
system; in other words, without an appropriate contact dynamics,
canonical and grand canonical ensembles would not be equivalent
\cite{Pradhan1_PRE2011, Mukamel_PRL2012}. The situation is 
analogous to that in equilibrium where equivalence of ensembles, 
a basic tenet of equilibrium thermodynamics, is ensured by the mass
exchange rates which satisfy detailed balance with respect to the
Boltzmann distribution. However in nonequilibrium, in the absence
of a priori knowledge of microscopic steady-state structure, the
intriguing questions, (a) whether there indeed exist a class of
exchange rates which could lead to the construction of a well-
defined nonequilibrium thermodynamics and (b) how the rates could 
be determined, are still unsettled.

Previous studies addressed some of these issues. However, the exact 
studies \citep{Bertin_PRL2006, Bertin_PRE2007} were mostly confined 
to a special class of models, called zero range processes. These 
models have product-measure or factorized 
steady state and therefore do not have any spatial correlations. In 
other studies, a class of lattice gas models with nonzero spatial 
correlations were considered \cite{SasaTasaki_JStatPhys2006, 
Pradhan_PRL2010, Dickman_PRE2014, Dickman2_PRE2014, 
Pradhan2_PRE2011} and, for some 
particular choice of mass exchange rates, zeroth law was found to 
be obeyed. However, the mass exchange rates, even in the limit of 
slow exchange, alters the fluctuation properties of the individual 
systems, leading to the breakdown of equivalence between canonical 
and grandcanonical ensembles.

In this paper, we formulate necessary and sufficient condition 
for which equilibrium thermodynamics can be consistently extended 
to {\it weakly interacting} nonequilibrium steady-state systems 
having {\it nonzero} spatial correlations. Under this 
condition, zeroth law is obeyed and ``equilibration'' between two 
systems (labeled by $\alpha=\1,\2$) in contact can be characterized 
by equalization of an intensive thermodynamic variable which is 
inherently associated with the respective isolated system. To 
obtain such a thermodynamic structure, we require the following 
condition: Mass exchange from one system to the other should occur 
weakly across the contact with the exchange rates satisfying \be
\frac{u_{\1 \2}(\mDelta)}{u_{\2 \1}(\mDelta)} = e^{-\Delta F},
\label{Balance_intro} \ee a reminiscent of detailed balance
condition in equilibrium. Here $u_{\alpha \alpha'}(\mDelta)$ is
the rate with which a mass of size $\mDelta$ is transferred from
system $\alpha$ to $\alpha'$, and $\Delta F$ is the change in a
nonequilibrium canonical free energy of the contact regions. In the 
limit of weak interaction between systems, the mass exchange rates 
are not necessarily small, but only that the mass exchange process 
do not affect the dynamics in the individual systems and spatial 
correlations between the systems vanishes. Note that 
Eq. \ref{Balance_intro} requires a free energy function inherent 
to individual isolated system to exist, which, we argue, is the 
case in a system having short-ranged spatial correlations. This 
free energy function can in principle be obtained from a 
fluctuation-response relation, analogous to fluctuation-dissipation 
theorems in equilibrium.

The notion of weak interaction is crucial to construct a 
well-defined nonequilibrium thermodynamics. Also in equilibrium, 
one implicitly assumes weak interaction where interaction energy 
between systems is taken to be vanishingly small so that bulk 
dynamics in an individual system remain unaffected by the other 
system which may be put in contact with the former. Likewise, 
weakly interacting nonequilibrium systems imply that dynamics in 
the individual systems remain unaffected even when two systems are 
kept in contact. The weak interaction limit, which essentially 
demands {\it vanishing} of correlations between two systems across 
the contact, is however not guaranteed by mere slow exchange 
of masses and {\it vice versa}. We demonstrate how the weak 
interaction limit can actually be achieved.

The organization of the paper is as follows. In section II.A, we 
discuss why an additivity property as in Eq. \ref{additivity1} is 
required for constructing a well-defined thermodynamic structure 
for nonequilibrium systems. In section II.B, we show that the 
coarse-grained balance condition (see Eq. \ref{Balance_intro}) on 
mass exchange rates ensures the desired additivity property. In 
section III, through various previously studied models and their 
variants,  we illustrate how the mass exchange rates can  
be explicitly constructed so that the balance condition Eq. 
\ref{Balance_intro} is satisfied. In section IV, we discuss
that generic mass exchange rates, even in the limit of slow 
exchange, leads to the breakdown of equivalence between canonical 
and grandcanonical ensembles. At the end, we summarize with a few 
concluding remarks and open issues.

\section{Theory}

\subsection{General considerations}

Let us consider two systems
$\alpha=\1,\2$ of size $V_\alpha,$ having mass variables ${\bf
m}_{\alpha}\equiv \{m_i \ge 0\}$ defined at the  sites $i \in
V_\alpha.$ Each of the systems, while not in contact with each
other (we refer to the situation as {\it canonical ensemble}), has
a nonequilibrium steady state distribution \be {\cal
P}_{\alpha}({\bf m}_{\alpha}) = \frac{\omega_{\alpha}({\bf
m}_{\alpha})}{W_{\alpha}(M_{\alpha}, V_{\alpha})} \delta\left(
M_{\alpha}- \sum_{i \in V_{\alpha}} m_i \right) \label{P_steady1}
\ee where $\omega_{\alpha}({\bf m}_{\alpha})$ is the steady-state
weight of a microscopic configuration ${\bf m}_{\alpha}$ and
$$W_{\alpha}(M_{\alpha}, V_{\alpha}) = \int d{{\bf m}_{\alpha}}
\omega_{\alpha} ({\bf m}_{\alpha}) \delta\left( M_{\alpha}-
\sum_{i \in V_{\alpha}} m_i \right), \label{Z1}$$ is the partition
sum ($\int d{{\bf m}_{\alpha}}$ implies integral over all mass 
variables $m_i$
with $i \in V_\alpha$). The delta function  ensures conservation
of mass   $M_{\alpha}=\sum_{i \in V_{\alpha}} m_i$, or mass density
$\rho_\alpha= M_{\alpha}/V_\alpha$, of individual systems. The
microscopic weight $\omega_{\alpha}({\bf m}_{\alpha})$  is the
time-independent solution of Master equation governing the
time evolution of the system in the configuration space of
${\bf m}_\alpha$ and in most cases is {\it not} known.
On the other hand,  when the systems  $\1$ and $\2$  are in
contact,  mass  exchange  from one system to the other at  the
contact region breaks conservation of $M_{\1,\2}$ whereas the
total mass $M=M_{\1} + M_{\2}$ of the combined system remains
conserved. We refer this situation as a {\it grand canonical
ensemble}.

To have a consistent thermodynamic structure, it is necessary that
individual systems themselves have well defined canonical free 
energy functions, $F_{\1,\2}$ for systems $\alpha = \1$ or $\2$. 
Moreover, this free energy function should not change due to the 
contact between the two systems. That is,
free energy of the combined system $F = F_\1+F_\2$ is obtained by
adding the corresponding canonical free energies of the individual
systems and the macrostate, or the maximum probable state, is
obtained by minimizing the total free energy function. This
additivity property has the following immediate
consequences: (i) Equalization of an intensive thermodynamic
variable, (ii) a fluctuation-response relation and (iii) zeroth
law; all of them follows from standard statistical mechanics
\cite{Kardar}.

\begin{figure}
\begin{center}
\leavevmode
\includegraphics[width=8.5cm,angle=0]{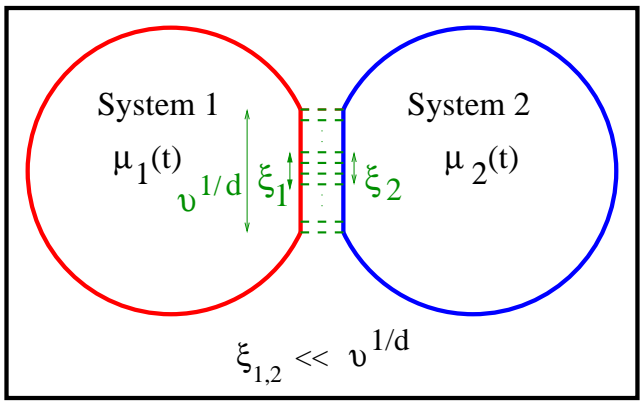}
\caption{(Color online) Schematic representation: ``Equilibration'' of two 
steady-state systems in contact. Intensive thermodynamic variables 
$\mu_1(t)$ and $\mu_2(t)$, chemical potentials of systems $1$ and 
$2$ at time $t$, eventually equalize in the steady state, 
$\mu_1(t=\infty)=\mu_2(t=\infty)$. The size of the contact region 
$v^{1/d}$, $v$ the volume of the contact region in $d$ dimension, 
is much larger than the individual correlation length $\xi_\alpha$.  }
\label{Diagram}
\end{center}
\end{figure}

First we discuss the macroscopic properties of systems in
canonical ensemble and how a free energy function can be defined
consistently for nonequilibrium systems. We consider an individual
system $\alpha$ divided into two subsystems, each of which being
much larger than  spatial  correlation length $\xi_\alpha$ and
total mass $M_\alpha$ being conserved. As subsystems much larger
than the correlation lengths would be statistically independent in
the thermodynamic  limit, the  steady state subsystem mass
distribution can be written as product of some weight factors which 
depend only on mass of the individual subsystem  \cite{Eyink1996, 
Bertin_PRL2006}. Thus, when $\xi_{\1,\2} \ll v \ll V_{\1,\2}$, we 
could view each individual system $\alpha$ composed of two 
statistically independent (apart from the constraint of total mass 
conservation provided by a delta function) macroscopically large 
subsystems - contact region (of size $v$ and mass $M_\alpha^c$) and 
the rest, i.e., the bulk (of size
$V_\alpha -v$ and mass $M_\alpha^b= M_\alpha - M_\alpha^c)$ -
whose steady-state weights are factorized, i.e., product of two
coarse-grained weights, as reflected in the partition sum \be
W_{\alpha}(M_{\alpha}, V_{\alpha}) \simeq \int
dM^c_{\alpha}W_{\alpha}(M_\alpha -M^c_{\alpha})
W_{\alpha}(M^c_{\alpha}). \label{factorw} \ee Or equivalently, the
joint probability distribution of subsystem masses will have a
factorized form \bea P(M_\alpha^c, M_\alpha^b) \simeq
\frac{W_\alpha(M_\alpha^c) W_\alpha(M_\alpha^b)}{W_\alpha(M_\alpha, 
V_\alpha)} \delta \left( M_\alpha - M_\alpha^c - M_\alpha^b \right) 
\nonumber \\
= \frac{e^{-[F_\alpha(M_\alpha^c) + F_\alpha(M_\alpha^b)]}}{e^{-
F_\alpha(M_\alpha)}} \delta \left( M_\alpha - M_\alpha^c - 
M_\alpha^b \right), \label{factorw1} \eea which is maximized to 
obtain macrostate of  the systems (i.e., the maximum probable 
state). These considerations immediately lead to the existence of a
canonical free energy $F_\alpha \equiv - \ln W_\alpha$ in the
steady state. The total steady-state free energy 
$F_\alpha(M_\alpha, V_\alpha)$
of the two subsystems is additive and is obtained by minimizing
the sum of free energy of the  bulk (of volume $V-v$) and that of
the contact region (of volume  $v$), \be F_\alpha(M_\alpha, V_\alpha) = {\rm
inf}_{M_\alpha^ c} [F_\alpha(M_\alpha^c, v) + F_\alpha(M_\alpha -
M_\alpha^c, V_\alpha - v )]. \ee The additivity property in Eq. 
\ref{factorw1} and the above minimization of total free energy
implies existence of an intensive thermodynamic variable, called 
chemical potential, \be \mu_\alpha(\rho_\alpha) = \frac{\partial 
F_\alpha}{\partial M_\alpha} = \frac{\partial f_\alpha}{\partial 
\rho_\alpha} \label{mu1} \ee which takes the same value for
any subsystems (macroscopically large). In the above equation, we 
have defined a nonequilibrium free energy density function 
$f_\alpha (\rho_\alpha) = F_\alpha/V_\alpha$.

Note that the nonequilibrium free energy function is defined in 
such a way that the principle of free energy minimization 
automatically holds. Interestingly,  for a steady-state system 
having a conserved mass, this free energy function as well as 
chemical potential can be calculated from subsystem mass 
fluctuations (as illustrated later in various models) and 
therefore has practical importance, e.g., describing phase 
coexistence \cite{Pradhan2_PRE2011, Arghya}, etc.

We next consider grandcanonical ensemble - a situation where mass
exchange takes place between two systems through contact regions
(see Fig. \ref{Diagram}) each with  volume $v$ (taken same for
both systems for simplicity) which is much larger than finite
spatial correlation  length $\xi_{\alpha}$ but otherwise {\it
arbitrary}. We demand  that the canonical description where $M_\1$
and $M_2$  are individually conserved,  must   be equivalent to
the grand canonical  ensemble  where  only  total mass $M=M_{\1} +
M_{\2}$ is conserved. That is, the microscopic weight of the
combined system must be a product of the individual canonical
microscopic weights and therefore the probability of a microscopic
configuration of the combined system should be given by \be {\cal
P}({\bf m}_{\1}, {\bf m}_{\2}) = \frac{\omega_{\1}({\bf m}_{\1})
\omega_{2}({\bf m}_{2})}{W(M)} \delta\left(M- M_\1 - M_\2 \right),
\label{P_steady2} \ee with the the partition sum of the combined
system being $$W (M, V) = \int dM_{\1} W_{\1}(M_1, V_1)
W_{\2}(M-M_1, V_2).$$ So the the joint distribution of individual 
system masses is also factorized and can be written as the product 
of the individual canonical weights, \bea P(M_{\1}, M_{\2}) = 
\frac{W_\1(M_\1,V_\1) W_\2 (M_\2,V_\2) }{W(M,V)} \nonumber \\
\times \delta\left(M- M_\1 - M_\2 \right), \label{additivity1}
\eea and thus additivity is ensured for the combined systems. 
That is, total free energy $F(M,V) \equiv - \ln W(M,V)$ of the 
combined system in the steady state is given by $$ F(M,V) = 
\inf_{M_\1} [F_\1(M_\1,V_\1) + F_\2(M-M_\1,V_\2)],$$ which is the 
sum of individual canonical free energies. This implies that the 
chemical potential equalizes upon contact, i.e., $\mu_\1 (\rho_1) = 
\mu_\2 (\rho_\2)$.

\subsection{Proof of the balance condition}

Now we show how, in the weak interaction limit, the 
balance condition in Eq. \ref{Balance_intro} ensures additivity
property in Eq. \ref{additivity1} - the main result of this
paper. Let mass exchange occur at the contact with rate $u_{\alpha
\alpha'}(\mDelta)$  where a mass $\mDelta$ is transferred from
system $\alpha$ to $\alpha'$. The rate may depend on both the mass
values at the two contact regions (the mass dependence not 
explicitly shown in $u_{\alpha
\alpha'}$). Mass conservation in the individual systems is then
broken in this process ($M_\alpha \to  M_\alpha -\mDelta$ and
$M_{\alpha'} \to  M_{\alpha'}+\mDelta$), generating a  mass flow.
To attain stationarity, average  mass current $J_{\1 \2}(\mDelta)$
generated  by all possible microscopic exchanges corresponding the 
rates $u_{\1 \2}$, where the chipped off mass $\mDelta$ flows from 
system $\1$ to $\2$, must be  balanced by  the reverse current 
$J_{\2 \1}(\mDelta)$. Note that, though the net steady-state 
current $|J_{\alpha \alpha'}(\mDelta) - J_{\alpha' \alpha}
(\mDelta)|$ from one system to the other (across the contact) is 
exactly zero, the individual systems can still be far away from 
equilibrium and can have nonzero steady-state mass currents in the 
bulk.

Since the total mass $M=M_{\1} + M_{\2}$ of the combined system is 
conserved, the current balance condition $J_{\1 \2} (\mDelta) = 
J_{\2 \1}(\mDelta)$ can be written, using only one of the mass
variables, say $M_{\1}$,  as \bea P(M_{\1}, M-M_{\1}) U_{\1
\2}(M_\1,\mDelta) = P( M_{\1}-\mDelta, M-M_{\1}+\mDelta) \nonumber
\\ \times U_{\2 \1}(M-M_\1+\mDelta, \mDelta). \label{recursion} 
\eea 
Here $U_{\alpha
\alpha'}(x,y) $  is an   effective  rate with which mass $y$ is
transferred from system $\alpha$, having mass $x$, to $\alpha'$.
The current balance, along with Eq. \ref{additivity1}, gives \be
\frac{U_{\1 \2}(M_\1,\mDelta)}{ U_{\2 \1}(M-M_\1+\mDelta,
\mDelta)}  =  e^{-\Delta F} \label{balance-condition1} \ee where
$\Delta F = \sum_{\alpha=1}^2 (F_{\alpha}|_{final} -
F_{\alpha}|_{initial}$) difference in free energy of the combined
system. Or equivalently, we write the above ratio of effective 
exchange rates as \bea \frac{U_{\1 \2}(M_\1,\mDelta)} { U_{\2 \1}
(M-M_\1+\mDelta, \mDelta)} = e^{(\mu_{\alpha} - \mu_{\alpha'}) 
\mDelta}, \label{balance-condition1A} \eea
where $\mu_{\alpha} = \partial F_{\alpha}/\partial M_{\alpha}$ is 
a nonequilibrium chemical potential (see Eq. \ref{mu1}) which is 
inherently associated with the individual system $\alpha$.

Next we obtain a condition on the actual microscopic exchange
rate $u_{\1\2}(\mDelta)$. We first use the expression of current
$J_{\alpha \alpha'}(\mDelta)=\langle u_{\alpha \alpha'} \rangle$
as the average mass transfer rate from system $\alpha$ to
$\alpha'$ and write $J_{\1 \2}(\mDelta)=\int \int d{\bf
m}_\1 d{\bf m}_\2 {\cal P} ({\bf m}_{\1}, {\bf m}_{\2}) u_{\1 \2}
(\mDelta)$ as given below
\bea
J_{\1 \2}(\mDelta) = \left[ \prod_{\alpha=1}^2 \int d{\bf m}_\alpha
\right] {\cal P} ({\bf m}_{\1}, {\bf m}_{\2}) u_{\1 \2}(\mDelta) 
\delta(M - \sum_{\alpha=1}^2 M_{\alpha}) 
\nonumber \\ 
= \frac{1}{W(M,V)} \int \int d{\bf m}_{\1} d{\bf 
m}_{\2}  \omega_{\1}({\bf m}_{\1}) \omega_{\2}({\bf m}_{\2}) 
u_{\1 \2}(\mDelta) 
\nonumber \\
\times \delta\left( M_{\1}- \sum_{i \in V_{\1}} m_i \right) \delta\left( M_{\2} - \sum_{i \in V_{\2}} m_i \right)
\delta(M - \sum_{\alpha=1}^2 M_{\alpha}) 
\nonumber \\
\simeq \frac{1}{W(M,V)} \int \int d{M^c_{\1}} d{M^{c}_{\2}}  
u_{\1 \2} W_{\1}(M^c_{\1}) W_{\2}(M^{c}_{\2})
\nonumber \\
\times  W_{\1}(M_{\1}-M^c_{\1}, V_{\1} - v) W_{\2}(M_{\2} - M^c_{\2}, V_{\2} - v). \mbox{~~}
\eea
In the last step, we inserted an identity $\int d{M^c_{\alpha}} 
\delta ( M^c_{\alpha} - \sum_{i \in v} m_i) = 1$, where $v$ being 
denoted here as the contact region in system $\alpha$, and then 
used the factorization property,
\bea
\int d{{\bf m}_{\alpha}} w_{\alpha}({\bf m}_{\alpha})  \delta\left( 
M_{\alpha}- \sum_{i \in V_{\alpha}} m_i \right) \delta 
\left( M^c_{\alpha} - \sum_{i \in v} m_i \right) \nonumber \\
\simeq  W_{\alpha}(M^c_{\alpha}, v) W_{\alpha}(M_{\alpha}-M^c_{\alpha}, V_{\alpha} - v), \nonumber
\eea as in Eq. \ref{factorw}.
As demonstrated later in various models in section III, the above 
factorization property is expected to be valid when the 
size of the contact region is much larger than the spatial 
correlation length $\xi_\alpha$ in system $\alpha$, i.e., when  
$v \gg (\xi_\alpha)^d$ in $d$ dimension. Then, after some 
straightforward manipulations, we write $U_{\1 \2} (M_\1, 
\mDelta)=J_{\1 \2}(\mDelta)/ P(M_1, M_2)$ as \bea
U_{\1 \2}(M_\1, \mDelta) = \int_{\mDelta} \int_0 d{M_\1^c}
d{M_\2^c} u_{\1 \2}(\mDelta) \prod_{\alpha=1}^2 \frac{  W_{\alpha}
(M_{\alpha}^c) e^{\mu_{\alpha} M_{\alpha}^c}} { {\cal Z}_{\alpha}}, 
\mbox{~~} \label{effective-rate1} \eea 
by using Eq. \ref{additivity1} and using the following equality
$$W_\alpha(M^c_\alpha) \frac{W_\alpha(M_\alpha - M^c_\alpha, 
V_\alpha-v)}{W_\alpha(M_\alpha, V_\alpha)} = 
\frac{W_\alpha(M^c_\alpha) e^{\mu_\alpha M^c_\alpha}}{{\cal 
Z}_\alpha}$$ where ${\cal Z}_\alpha = \int dM_\alpha^c  ~ 
W_{\alpha}(M^c_{\alpha}) ~ e^{\mu_\alpha M_\alpha^c}$. Similarly, 
the effective reverse exchange rate, corresponding to the 
transition $\{M^c_{\1}-\mDelta, M^c_{\2}+\mDelta \} \rightarrow 
\{M^c_{\1}, M^c_{\2} \}$, can be written as \bea 
U_{\2 \1}(M-M_\1+\mDelta, \mDelta) = e^{(\mu_{\2} - \mu_{\1}) \mDelta} \int_{\mDelta}  \int_0 d{M^c_{\1}} d{M^c_{\2}}  u_{\2 \1}(\mDelta)
\nonumber \\
\times  \frac{W_{\1}(M^c_{\1}-\mDelta) e^{\mu_{\1} M^c_{\1}}}{{\cal  Z}_{1}} \frac{W_{\2}(M^c_{\2}+\mDelta) e^{\mu_{\2} M^c_{\2}}}
{{\cal Z}_2}.~~\label{effective-rate2} 
\eea
Now, substituting Eqs. \ref{effective-rate1} and 
\ref{effective-rate2} in Eq. \ref{balance-condition1A} and then by 
equating the integrals which is valid for any functional form of 
weight factor $W_{\alpha} (m)$, we get the desired balance 
condition as in Eq. \ref{Balance_intro}, \bea \frac{u_{\1 \2}}
{u_{\2 \1}} = \frac{W_{\1}(M^{c}_{\1}- \mDelta)}{W_{\1}
(M^{c}_{\1})} \frac{W_{\2}(M^{c}_{\2}+\mDelta)}{W_{\2}(M^{c}_{\2})} 
= e^{- \Delta F^c} = e^{- \Delta F}. ~~\label{effective-rate3} \eea 
In the last step, we used the free energy of the contact region 
$F_\alpha^c(M_\alpha^c) = -\ln W_\alpha(M^c_\alpha)$ and equate 
the change in free energy at the contact $\Delta F^c = 
\sum_{\alpha=1}^2 \Delta F_\alpha^c$ to  the change in total free 
energy of the combined system $\Delta F$. This is so since the 
total free energy $F = \sum_{\alpha=1}^2 (F^c_\alpha + F^b_\alpha)$ 
can be written as a sum of bulk free energy $F^b_\alpha$ and 
contact free energy $F^c_\alpha$ where $\Delta F^b_\alpha=0$ (i.e., 
changes occur only at the contact regions). Note that the balance 
condition holds only at the contact regions for mass transfer from 
one system to the other. However, there is no detailed balancing in 
the bulk, except when both the systems are in equilibrium.

The balance condition in Eq. \ref{effective-rate3} is necessary and 
sufficient to ensure that the steady state has the required product 
form as in Eq. \ref{P_steady2}. This is because any contact 
dynamics which is constrained by the balance condition in
Eq. \ref{effective-rate3} indeed satisfies Master equation in the 
steady state as the mass-current balance condition $J_{\1 \2}
(\mDelta) = J_{\2 \1}(\mDelta)$, used for deriving the balance 
condition Eq. \ref{Balance_intro}, is nothing but the balancing of 
configuration-space current occurring due to exchange of masses.  
This completes the proof.

Note that Eq. \ref{effective-rate3} does not uniquely specify the 
contact dynamics (CD); two simple choices which we discuss in this 
paper are given below,
\bea
{\rm CD~I}&:& u_{\alpha \alpha'} =  u_0 p(\mDelta)
\frac{W_{\alpha}^c(M^{c}_{\alpha} -
\mDelta)}{W_{\alpha}^c(M^{c}_{\alpha})},
\label{rate1} \\
{\rm CD~II}&:& u_{\alpha \alpha'}=   u_0 p(\mDelta) {\rm Min}\{1,
e^{-\Delta F}\}, \label{rate2} 
\eea 
where $u_0$ an arbitrary constant (not necessarily small) and 
$p(\mDelta)$ is  a probability that mass $\mDelta$ is chosen for 
exchange. Note that the limit $u_0 \rightarrow 0$ implies slow 
exchange of masses. The case with $u_0=0$ implies no exchange of 
masses, i.e., the systems are kept isolated. The resemblance 
between the rate in Eq. \ref{rate2} and the familiar Metropolis 
rate is indeed striking. In equilibrium, Eq. \ref{effective-rate3} 
reduces to the condition of detailed balance, albeit on a coarse-
grained level. A similar notion of coarse-grained detailed balance 
was previously envisaged in \cite{Bertin_PRE2007}, though in the 
context of zero range processes which do not have any spatial 
correlations.

What still remains to be done is to explicitly specify the 
exchange rates satisfying Eq. \ref{effective-rate3}. This requires 
calculation of the subsystem weight factor $W_{\alpha}(m,v)$ in a 
particular system of interest, which can be done following Ref. 
\cite{PRL2014}. Note that the Laplace transform $\tilde{W}_{\alpha}
(s,v) = \int_0^{\infty} W_{\alpha} (M^c_{\alpha}) \exp(-s 
M^c_{\alpha}) dM^c_{\alpha}$ of the subsystem weight factor can be 
written in terms of the Laplace transform $\tilde{W}_\alpha(s, 
V_{\alpha}) = \int_0^{\infty} W_\alpha(M_{\alpha}, V_{\alpha}) 
\exp(- s M_{\alpha}) dM_{\alpha}$ of the individual canonical 
partition sum $W_\alpha(M_{\alpha}, V_{\alpha})$ as
\be
\tilde{W}_{\alpha}(s,v) = \left[ \tilde{W}_\alpha(s, V_{\alpha}) 
\right]^{v/V_\alpha},
\label{LT-W}
\ee
in the limit $V_\alpha \gg v \gg \xi_\alpha^d$ (in $d$ dimensions). 
The partition sum $W_\alpha(M_{\alpha}, V_{\alpha})$ can be 
calculated, as follows, from a canonical fluctuation-response relation. The subsystem mass fluctuation when calculated in 
canonical ensemble with $u_0=0$ is related to the change in density 
$\rho_\alpha$ in response to the change in chemical potential 
$\mu_\alpha$ (as in Eq. \ref{mu1}) as given below 
\be 
\frac{d\rho_{\alpha}}{d\mu_\alpha} = {\psi_\alpha 
(\rho_\alpha)}, \label{FR1} 
\ee 
where, for subsystem volume $v \gg \xi_\alpha^d$, the function 
$\psi_\alpha(\rho_\alpha) = \sigma^2_v/v$ with variance of 
subsystem mass $\sigma_v^2 = \langle {(M_\alpha^c)}^2 \rangle -v^2 
\rho_\alpha^2$. The variance of subsystem mass in system $\alpha$ 
can be calculated from the knowledge of correlation function 
$c_\alpha(r)$ as $\sigma^2_v \simeq v \sum_{r=-\infty}^{r=\infty} 
c_{\alpha}(r)$ where $c_\alpha(r) = \langle m_i m_{i+r} \rangle 
-\rho_\alpha^2$ is the two-point correlation between masses at 
sites $i$ and $i+r$ \cite{PRL2014}. We assumed here that the 
correlation function $c_{\alpha} (r)$ is short-ranged or 
sufficiently rapidly decaying function so that it is integrable, 
which is usually the case when there is no long-ranged correlations 
in the systems. Therefore, once the functional dependence of 
$\psi_\alpha(\rho_\alpha)$ on the respective density is known, the 
partition sum for individual system $W_\alpha(M_\alpha, V_\alpha)
= \exp[-V_\alpha f_\alpha(\rho_\alpha)]$, $f_\alpha(\rho_\alpha)$ 
being nonequilibrium free energy density, can be obtained by 
first integrating the fluctuation-response relation Eq. \ref{FR1}
w.r.t. density $\rho_{\alpha}$ and then integrating chemical 
potential as given in Eq. \ref{mu1}. Then the subsystem weight 
factor $W_{\alpha}(m)$ can be obtained, via inverse Laplace 
transform, from Eq. \ref{LT-W}.

We emphasize here that, even when the detailed microscopic weight 
$\omega_\alpha({\bf m}_\alpha)$ is not known, the subsystem weight 
factor $W_\alpha(m,v)$ can still be obtained, either analytically 
or numerically, from the subsystem mass fluctuations or 
equivalently from the two-point spatial correlation functions; this 
makes our formulation work both in theory and in practice.

\section{Models and Illustrations}

In this section, we illustrate our analytical results in  
nonequilibrium models studied extensively in the past as well as 
in their variants. For each of these models, we analytically 
obtain chemical potential $\mu(\rho_\alpha)$ and the weight factor 
$W_\alpha(m)$ when the system is isolated (i.e., $u_0=0$), and then
we explicitly construct the mass exchange rates $u_{\alpha 
\alpha'}$ so that they satisfy the balance condition Eq. 
\ref{effective-rate3}. Using these 
rates, we perform simulations (we use both the contact dynamics I 
and II). Our simulations demonstrate that, when two systems are 
kept in contact with unequal initial individual chemical 
potentials, they indeed ``equilibrate'' where the chemical 
potentials associated with the respective isolated systems equalize 
in the final steady state of the combined system.

\subsection{Zero Range Processes}

For completeness, we first consider zero range processes (ZRP) 
\cite{Haney2005} which have a factorized steady state (FSS). For
ZRP, a well-defined thermodynamic structure has been previously 
constructed \cite{Bertin_PRE2007}. Consider two systems $\alpha=\1,
\2$ where their steady-state weights
$$\omega_\alpha({\bf m}_\alpha) = \prod_{i\in V_\alpha}
h_\alpha(m_i)$$ are simply product of factors $h_\alpha(m_i)$,
function of only single-site mass variable. The
individual systems exactly satisfy Eq. \ref{factorw} with  weights
of contact region (volume $v$) and the rest of system (volume 
$V_\alpha-v$) being $W_\alpha^c = (f_\alpha)^{v}$ and $W_\alpha^b = 
(f_\alpha)^{V_\alpha-v}$, respectively. When mass exchange occurs 
either with rate CD I (Eq. \ref{rate1}) or with rate CD II (Eq. 
\ref{rate2}), it is easy to check that the joint distribution, 
which satisfies Master equation, is given by
$${\cal P}({\bf m}_1, {\bf m}_2) \propto \prod_\alpha \prod_{i\in 
V_\alpha} \exp[- f_\alpha(m_i)],$$ i.e., product of individual 
weight factors $\omega_\alpha ({\bf m}_\alpha)$ with 
$f_\alpha(m_i) = - \ln h_\alpha(m_i)$. Note that, for FSS, Eq. 
\ref{effective-rate3} indeed reduces to detailed balancing at the 
contact, as found in \cite{Bertin_PRE2007}.

\subsection{Finite Range Processes}

Now we consider a general situation - keeping in contact systems
having nonzero {\it spatial correlations}. To this end, we 
introduce a broad class of analytically tractable models, for 
simplicity in one dimension, where a particle (or mass of size 
$\mDelta$) is transferred stochastically from a site to one of its 
nearest-neighbours with rates depending on the discrete occupation 
numbers (or continuous mass variables) of $R$ neighbouring sites. 
These models are direct generalization of the zero range processes 
\cite{PRL2014, Amit} and are called here finite range processes, 
with range $R$. These finite range mass transport processes have a 
clusterwise factorized steady state  (CFSS) where each weight 
factor depends on the occupation numbers (or mass variables) $m_i$ 
($i \in R$) of a cluster of size $R$. We consider two  systems 
$\alpha=\1,\2$, for simplicity on two one dimensional periodic 
lattices of individual size $L_\alpha$, where each system having a 
CFSS of form $$ \omega_\alpha ({\bf m}_\alpha) = 
\prod_{i=1}^{L_\alpha} g_{\alpha} (m_i, m_{i+1}, \dots, 
m_{i+R})$$ where $g_\alpha$ a function of $R+1$ mass or occupation 
variables at consecutive $R+1$ sites. Clearly, $R=0$ corresponds to
the factorized steady state (FSS) as in ZRP. The CFSS could arise 
in a variety of mass transport processes where mass chipping rate 
in the bulk satisfies certain conditions, details of which will be 
provided elsewhere \cite{Amit}. Below, we consider only the 
{\it continuous} mass CFSS.

Unlike ZRP, the joint distribution of masses is not factorized on 
the single-site level as $g(m_i,m_{i+1}, \dots, m_{i+R})$ is 
function of masses at $R+1$ sites and therefore generates finite 
spatial correlations. In this paper, mainly due to analytical 
tractability, we consider a special form of $g_\alpha(m_i,m_{i+1}, 
\dots, m_{i+R})$ which is a homogeneous function,
\be
g_\alpha(\Gamma m_i, \Gamma m_{i+1}, \dots, \Gamma m_{i+R}) = 
\Gamma^{\delta} g_\alpha(m_i,m_{i+1},\dots, m_{i+R}) \label{Homo2}
\ee
with $\delta$ real. For this particular form, the two-point 
correlation function $c_\alpha(r)$ can be exactly calculated. By 
rescaling of mass variable $m_k = \rho_\alpha m'_k$ in the 
individual isolated system with density $\rho_\alpha$, 
correlation of masses $\langle m_i m_{i+r} \rangle$ at sites $i$ 
and $i+r$ can be written, as $\langle m_i m_{i+r} \rangle = A_\alpha(r) \rho_\alpha^2$  where 
\small
\be
A_\alpha(r) = \frac{ \prod_k \left[\int_0^{\infty} 
dm'_k g_\alpha^{(k)} (\{ m'_k \}_{_R}) \right] m'_i m'_{i+r} 
\delta \left(\sum_k m'_k - L_\alpha \right)}{\prod_k 
\left[\int_0^{\infty} dm'_k g_\alpha^{(k)} (\{ m'_k \}_{_R}) 
\right] \delta \left(\sum_k m'_k - L_\alpha 
\right)}, 
\ee
\normalsize
$g_\alpha^{(k)}(\{ m'_k \}_{_R}) \equiv \rho_\alpha^{-\delta} 
g_\alpha(m_k, m_{k+1} \dots, m_{k+R})$ \cite{PRL2014}. The function 
$A_\alpha (r)$ depends on relative distance $r$, but is independent 
of density $\rho_\alpha$, and can be exactly calculated 
using a transfer matrix method \cite{Amit}. Then, in an individual 
system $\alpha$, we obtain variance in a subsystem of size $v$ as 
$\sigma^2_v = v \rho_\alpha^2/\eta_\alpha$ with $\eta_\alpha^{-1} = \sum_{r=-\infty}^{\infty} [A_\alpha(r) - 1]$. 
Now the subsystem weight factor $W_\alpha(m)$ can be exactly 
calculated, using the method outlined in the end of section II.B,
to get a functional form of $W_\alpha(m) = m ^{v \eta_\alpha -1}$.

In the case of nonzero spatial correlations, by considering a 
system in a coarse-grained level, one can have physical insights 
into the role of the balance condition Eq. \ref{Balance_intro}.
Let us divide a system $\alpha$ into 
$\nu_{\alpha} = V_{\alpha}/v$ number of almost {\it statistically 
independent} subsystems of equal volume $v$ with subsystem masses 
labelled by  ${\bf\cal M}_\alpha \equiv \{{\cal M}_{\alpha,j}\}$, 
provided that the spatial correlation length $\xi_{\alpha}$ is much 
smaller than $v^{1/d}$ (in $d$ dimensions). Then the joint
probability distribution of the subsystem masses  of systems
$\alpha$ are factorized: $${\cal P}(\{{\bf \cal M}_{\1}, {\bf \cal
M}_{\2} \}) \propto \prod_\alpha \prod_{j \in V_\alpha}
\exp[-F^{(\alpha)} (\{{\cal M}_{\alpha,j}\})]$$ where free energy
$F_\alpha= - \sum_j \ln W_{\alpha}({\cal M}_{\alpha,j})$ of system
$\alpha$ is additive over the subsystems. Now let two such systems
$\1$ and $\2$ be kept in contact such that mass from one specific
subsystem  of $\1$ participate in a microscopic mass-exchange
dynamics with its adjacent subsystem  of $\2$ with rates
satisfying Eq. \ref{effective-rate3}. In a coarse-grained level,
as the subsystems could be considered as sites, the systems
effectively become a set of sites with an ``FSS'', where mass
exchange occurs between two adjacent sites (here subsystems) with
rates satisfying balance condition Eq. \ref{effective-rate3}, and
therefore the additivity property in Eq. \ref{additivity1} holds
exactly in the limit of large subsystem volume $v \gg \xi_{\1, 2}$.

Next, we discuss in detail a special case of the clusterwise 
factorized steady state with $R=1$.

\begin{figure*}
\begin{center}
\leavevmode
\includegraphics[width=18.0cm,angle=0]{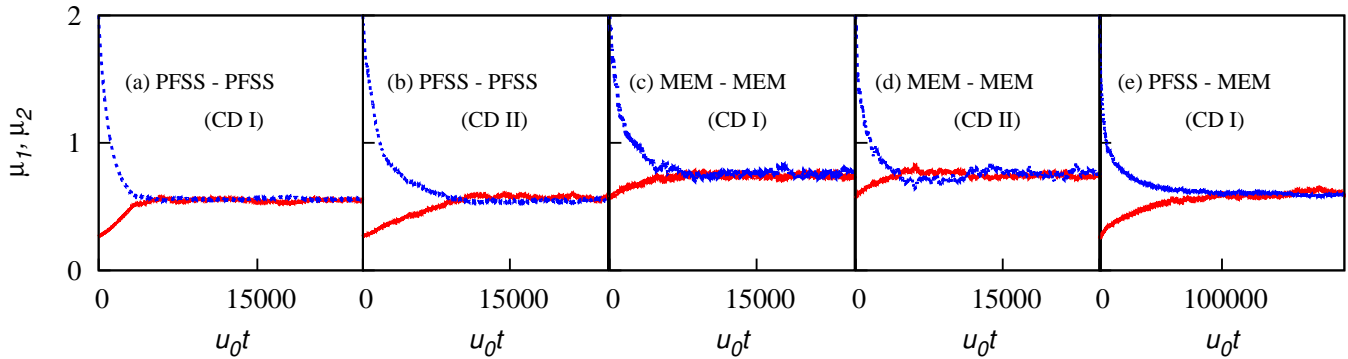}
\caption{(Color online) ``Equilibration'' of steady states in 
contact: In (a)-(e), chemical potentials $\mu_\1(t)$ and 
$\mu_\2(t)$ of systems $\1$ (red solid lines) and $\2$ (blue dotted 
lines) vs.  rescaled time
$u_0 t$. $\mu_\1$ and $\mu_\2$, initially chosen to be different,
eventually equalize. Densities ($\rho_\1$, $\rho_\2$) in the final
steady states  are respectively ($3.60$, $5.40$) in (b), ($3.57$,
$5.43$) in (c), ($5.31$, $2.69$) in (d), ($5.32, 2.68$) in (e),
and  ($3.32$, $6.68$) in (f). In all cases, $p(\mDelta) =
p_b(\mDelta) = \exp(-\mDelta)$, $u_0= 0.1$ and $v=10$ (except in
(d) and (e) where $v=1$).} \label{Equalization1}
\end{center}
\end{figure*}

\subsection{Pair Factorized Steady State (PFSS)}

To demonstrate that our results are valid even in the presence of
nonzero spatial correlations, we first consider two one-dimensional 
periodic lattices  of  $L_\alpha$ sites with continuous mass 
variable $m_i \ge 0$  at  sites  $i=1,2, \dots, L _\alpha.$ 
The following mass conserving dynamics in the bulk leads to a CFSS
with $R=1$, usually called pair factorized steady state (PFSS)
\cite{Evans_PRL2006}, where mass $\mDelta$ chosen
from a distribution $p_b(\mDelta)$ is chipped off from a site $i$
and transferred to its right neighbor with rate \bea
u_{\alpha}^b(\mDelta) = p_b(\mDelta) \frac{g_{\alpha}(m_{i-1}, m_i
- \mDelta)}{g_{\alpha}(m_{i-1}, m_i)}
\frac{g_{\alpha}(m_i-\mDelta, m_{i+1})}{g_{\alpha}(m_i, m_{i+1})},
\label{bulkrate_PFSS} \eea which  depends on the masses at the
departure site and its nearest neighbors, and on the chipped-off
mass $\mDelta$. Since, in this case, mass-transfer happens in only
one direction in the bulk, there are nonzero bulk currents present
in the individual systems. We consider homogeneous  $g_\alpha(x, y) 
= \Gamma^{-\delta} g_{\alpha}(\Gamma x, \Gamma y),$ for which
one can exactly calculate $\psi_\alpha(\rho_\alpha) = \rho_\alpha^2
/\eta_\alpha$ and values of $\eta_\alpha$ for various microscopic
parameters \cite{Amit}. Then following the method outlined in
section II.B, we analytically obtain $W_{\alpha}(m) = m^{v 
\eta_\alpha-1}$ and chemical potential
$\mu_{\alpha}=-\eta_{\alpha}/\rho_{\alpha}$ where  $\eta_\alpha$
depends on $\delta$. When  two such  systems are kept  in contact,
mass conservation in individual system is broken and both density
$\rho_\alpha(t)$ and corresponding chemical potential
$\mu_\alpha(t)$ evolve until a stationarity is reached where the
net mass current from one system to another vanishes and densities
are adjusted so that chemical potentials equalize. We simulate
using $g_\alpha(x,y) =(x^\delta+y^\delta+c x^\gamma
y^{\delta-\gamma})$ and allow the  two PFSS with $\eta_1=2$
($\delta=1$, $c=0$) and $\eta_2=3$ ($\delta=2$, $c=1$,
$\gamma=3/2$) to exchange mass  following CD I (and CD II in
different simulations) with  $u_0=0.1$, $p(\mDelta)=
p_b(\mDelta)=exp(-\mDelta)$ and $L_\1=L_\2 =1000$. The contact
volume $v=10$ is taken much larger than $\xi_\alpha$ which is
here only about a couple of lattice spacings. Simulations in Figs.
\ref{Equalization1}(a) and \ref{Equalization1}(b) demonstrate
that, starting from arbitrary initial densities, the combined
system reaches a stationary state where  $\mu_1 = \mu_\2.$

The equalization of an ITV, i.e., the above  mentioned chemical
potential, indeed implies zeroth law which we verify next for
three steady states having a PFSS: PFSS$1$ ($\delta=1$, $c=0$;
$\eta_1=2$), PFSS$2$ ($\delta=3$, $c=0$; $\eta_2=4$) and PFSS$3$
($\delta=2$, $c=1.0$, $\gamma=3/2$; $\eta_3=3$) with CD I. First,
PFSS$1$ with density $\rho_1 \simeq 3.60$ and PFSS$2$ with density
$\rho_2 \simeq 7.25$ are separately equilibrated with a third
system PFSS$3$ with density $\rho_3 \simeq 5.37$. Then, PFSS$1$
with density $\rho_1$ and PFSS$2$ with density $\rho_2$ are
brought into contact. The two resulting densities after
equilibration remain almost unchanged, confirming zeroth law. The
zeroth law can be similarly verified for CD II.

\subsection{Mass Exchange Models (MEM)}

There are numerous examples \cite{MajumdarPRL1998, KrugGarcia2000,
RajeshMajumdar2000, Zielen_JSP2002, Mohanty_JSTAT2012}, where
nonequilibrium processes with a conserved mass show short-ranged
spatial correlations, but the exact steady-state structures are
not known. How does one find a contact dynamics which ensures Eq.
\ref{additivity1} in these cases? We address the question in a
class of  widely studied nonequilibrium mass transport processes
\cite{KMP1982, CC2000, Matthes-Toscani_JStatPhys2008,
Yakovenko_RMP2009}, as another demonstration of how our
formulation can  be implemented in practice. In these models, we
call them mass exchange models (MEM), in one dimension the
continuous masses $m_i \ge 0$ and $m_{i+1} \ge 0$ at randomly
chosen nearest neighbors $i$ and $i+1$ respectively are updated
from time $t$ to $t+dt$ as $$ m_i(t+dt) = \lambda_{\alpha} m_i(t)
+ r(1-\lambda_{\alpha}) m_{sum}(t),$$
 $$m_{i+1}(t+dt) = \lambda_{\alpha} m_{i+1}(t) +
(1-r)(1-\lambda_{\alpha}) m_{sum}(t),$$ where, $m_{sum}=m_i+m_{i+1}$
is the sum of nearest neighbor  masses, $r$ is a random number
uniformly distributed in $[0,1]$, and $0<\lambda_{\alpha} <1$ a
model dependent parameter. As the spatial correlations are nonzero
but very small,  the subsystem weight factor in the steady states
of individual systems can be obtained, to a very
good approximation, as $W_{\alpha}(m) = m^{v \eta_{\alpha}-1}$
with $\eta_{\alpha}=(1+2\lambda_{\alpha})/(1-\lambda_{\alpha})$
\cite{PRL2014}. In panels (c) and (d) of Fig. \ref{Equalization1},
we observe equalization of chemical potentials $\mu_\1(t) = -
\eta_\1/\rho_\1(t)$ and $\mu_\2(t) = - \eta_\2/\rho_\2(t)$
(respective ITV in this case) of systems $1$ and $2$ respectively
for both contact dynamics I and II and for $u_0=0.1$, $v=1$,
$L_\1=L_\2=100$ and $p(\mDelta)=p_b(\mDelta)=\exp(-\mDelta)$. The
zeroth law  can be readily verified for MEM as done in the case of
PFSS.

There is no  particular difficulty when systems having different
kind of bulk dynamics are  in contact; equilibration occurs as
long  as there is a common conserved quantity which is exchanged
following Eq. \ref{effective-rate3}. We demonstrate this in panel
(e) of Fig. \ref{Equalization1}, taking two  systems, PFSS and
MEM,  in contact  where mass exchange dynamics at the contact is
governed by CD I. In this case, chemical potentials $\mu_\1(t)$
and $\mu_\2(t)$ eventually equalize and zeroth law follows.

\section{Equivalence of ensembles}

In the previous section, we have demonstrated that, when two 
nonequilibrium systems with short-ranged correlation are allowed 
to exchange a conserved quantity  following a contact dynamics 
conditioned by Eq. \ref{Balance_intro}, they indeed evolve to a 
stationary state where an intensive thermodynamic variable (ITV), 
which is inherently associated with the respective isolated system, 
equalizes. In this thermodynamic construction, zeroth law is obeyed 
and, at the same time, equivalence of ensembles is also maintained. 
However, note that mere equalization of an intensive thermodynamic 
variable, or validity of zeroth law, does not guarantee the balance 
condition Eq. \ref{Balance_intro} and is not enough to construct a 
consistent nonequilibrium thermodynamics. This is because the ITV 
which equalizes for systems in grandcanonical ensemble is not 
necessarily the ITV defined (using additivity property Eq. 
\ref{factorw1}) for individual isolated systems in canonical 
ensemble. To construct a well-defined thermodynamic structure, one 
must ensure that these two ITVs are indeed the same. That is, one 
requires that the combined system (grandcanonical ensemble) is  
statistically equivalent to the individual isolated systems 
(canonical ensemble).

The requirement of ensemble equivalence, which  essentially demands 
that the contact dynamics  must not alter the fluctuation 
properties in the individual systems, is nothing special in 
nonequilibrium  scenario; it has been an essential ingredient in 
constructing equilibrium thermodynamics. The proposed balance 
condition Eq. \ref{Balance_intro} precisely ensures these two 
aspects - in one hand, it ensures equalization of an intensive 
thermodynamic variable and, on the other hand, it guarantees 
ensemble equivalence.

Note that, unlike in equilibrium, when two nonequilibrium systems 
are brought into contact, the final steady state of the combined 
system depends, in general, on the absolute values of mass exchange 
rates, even if the ratio between forward and reverse exchange rates 
remains unchanged. In these cases too, in the limit of slow 
mass exchange ($u_0 \rightarrow 0$) and weak interaction, there 
could exist an ITV which equalizes upon contact. However, in spite 
of the equalization of an ITV, as we illustrate in the following 
subsections, the mass exchange rates which do not satisfy the 
balance condition Eq. \ref{Balance_intro} lead to the breakdown 
of ensemble equivalence. That is, mass fluctuation in the isolated 
systems can be different from that in the combined system and, in 
that case, an equilibrium-like thermodynamic structure cannot be 
formulated.

In the examples given below, we consider weakly interacting lattice 
gases (driven and nondriven both) which exchange masses 
infinitesimally slowly, i.e.,  $u_0 \rightarrow 0$. The limit of 
slow exchange is useful in exactly calculating the mass 
fluctuations as the inhomogeneities which could occur in the 
contact regions of the individual systems is avoided and the weak 
interaction limit is also achieved.

\subsection{Lattice gases}

We start with $d$-dimensional lattice gases with interacting 
particles, obeying hardcore exclusion (at most one particle at a 
site). We consider periodic boundaries, though the following 
analysis can be 
straightforwardly extended to other boundary conditions (e.g., 
reflecting boundary, discussed in the case of nearest-neighbour-
exclusion lattice gases in section IV.B). {\it Internal dynamics}: 
Particles hop, from one site to its nearest neighbour, inside the 
individual systems according to some specified rates, e.g., rates 
satisfying local detailed balance \cite{KLS} with respect to the 
Boltzmann distribution $\sim \exp(-\beta E)$ where $\beta$ inverse 
temperature, $E_\alpha$ energy function of system $\alpha$ and 
$E=E_\1+E_\2$ total energy. {\it Mass exchange or contact 
dynamics}: The rate with which a particle at the contact region 
(which could be localized, even a point or single-site contact or 
global contact) jumps from system $\alpha$ to $\alpha'$, provided 
the contact site in $\alpha$ is occupied and the contact site in 
$\alpha'$ is unoccupied, is simply a constant $u_0 p_\alpha$. There 
is no additional constraint on these rates except that $u_0 
\rightarrow 0$ so that particle exchange occurs very slowly.

Since the transition rates overall do not satisfy detailed balance, 
the probability of a microscopic configuration of the combined 
system is not given by the Boltzmann distribution $\sim \exp(-\beta 
E)$. Note that the particle hopping rates inside the individual 
systems remain the same irrespective of two systems being in 
contact or not, which is necessary in realizing the weak 
interaction limit (which, for a finite $u_0$, is however {\it not} 
sufficient).

The joint probability distribution $P(M_\1, M_\2)$ of particle 
numbers $M_\1$ and $M_\2$ of individual systems, i.e., the
large deviation function governing mass or particle-number 
fluctuations, can be exactly calculated using the general recursion 
relation Eq. \ref{recursion}, with setting $\mDelta=1$ (i.e., one-
particle transfer at a time), as
\bea
P(M_\1, M_\2) = P(0,M) \prod_{M_\1} \frac{U_{\2 \1}(M_\2+1, 1)}{U_{\1 \2}(M_\1, 1)} \delta(M-\sum_{\alpha=1}^2 M_{\alpha}) \nonumber 
\\
= P(0,M) \left[ e^{\sum_{M_\1=0}^{M_\1} (\ln U_{\2 \1} - \ln U_{\1 \2})} \right] \delta(M-\sum_{\alpha=1}^2 M_{\alpha}). \label{LG1} ~~~
\eea
Now writing the effective mass exchange rates $U_{\1 \2}=u_0 p_\1 
\rho_\1 (1-\rho_2)$ and $U_{\2 \1}=u_0 p_2 \rho_\2 (1-\rho_1)$ and 
integrating over densities, the joint mass distribution can be 
exactly written in the form as given below, 
\bea
P(M_\1, M_\2) \propto e^{-\left[ V_\1 f_\1 + V_\2 f_\2 \right]}  \delta(M-V_\1 \rho_\1 - V_\2 \rho_\2) \label{LG2}
\eea
where free energy densities $f_\1 = \int_0^{\rho_\1} \mu_\1 
d\rho_\1$ and $f_\2=\int_{M/V_1}^{\rho_\2} \mu_\2 d\rho_\2$ with
chemical potential given by \be \mu_\alpha(\rho_\alpha)=\ln 
p_\alpha + \ln \frac{\rho_\alpha}{1-\rho_\alpha}. 
\label{mu_hardcore} \ee   
It is somewhat surprising that the joint mass distribution, as in 
Eq. \ref{LG1} or \ref{LG2}, is actually {\it independent} of the 
internal dynamics in each systems. Moreover, the above free energy 
and chemical potential are nothing but those of a noninteracting 
hardcore lattice gas. The macrostate, or the maximum probable 
state, of the combined system with final steady-state densities in 
the individual systems can be obtained by minimizing the total free 
energy $F=V_\1 f_\1 + V_2 f_\2$, with the constraint $V_\1 \rho_\1 
+ V_\2 \rho_\2={\rm constant}$. In other words, there exists an 
intensive thermodynamic variable, we call chemical potential, which 
indeed equalizes upon contact, i.e., $\mu_\1 (\rho_\1)= \mu_\2 
(\rho_\2)$. The equalization of chemical potential essentially 
signifies the steady-state current balance between two systems 
across the contact as encoded in Eq. \ref{recursion} and moreover 
this immediately leads to zeroth law under this particular contact 
dynamics.

However, in the above construction, clearly there is breakdown of 
equivalence between canonical and grandcanonical ensembles and, 
therefore thermodynamically, the construction is not well-defined. 
Note that, in this 
case, the free energy and chemical potential are not the same as 
those defined in canonical ensemble (see Eq. \ref{factorw1}) when 
$u_0=0$. In fact, in the canonical ensemble, subsystem 
particle-number fluctuation in individual systems can have 
nontrivial properties due to the presence of inter-particle 
interactions. But, with the above contact dynamics, the 
particle-number fluctuation in the grand-canonical ensemble is 
governed by a chemical potential of a noninteracting hardcore 
lattice gas (see Eq. \ref{mu_hardcore}), which is so in spite of
the presence of inter-particle interaction in the individual 
systems. The origin of the discrepancy in fluctuations in the 
two cases with $u_0=0$ and $u_0 \rightarrow 0$ lies in the fact 
that mass exchange rates do not satisfy the balance condition Eq. 
\ref{Balance_intro}, which drastically changes the fluctuation 
properties of the systems in grandcanonical ensembles. That is, 
unless the balance condition Eq. \ref{Balance_intro} is satisfied 
by the mass exchange rates, the cases with $u_0 = 0$ and $u_0 
\rightarrow  0$ are always different.

For example, inequivalence of ensembles arises in the previous 
studies \cite{SasaTasaki_JStatPhys2006, Pradhan_PRL2010, 
Pradhan2_PRE2011} where two driven lattice gases are allowed to 
exchange particles with some exchange rates, which were chosen on 
an {\it ad hoc} basis.  To be specific, let us consider the systems 
studied in  \cite{Pradhan2_PRE2011}, where two lattice gases - a 
nondriven lattice gas $\1$ and a driven lattice gas $\2$ (Katz-
Lebowitz-Spohn model \cite{KLS}), are kept in contact. Particle 
hopping rates in the bulk as well as the particle exchange rates 
across the contact both satisfy a local detailed balance 
\cite{KLS}. In the limit of slow mass exchange, the ratio of the 
effective transition rates was found, to a good approximation, to 
be \cite{Pradhan2_PRE2011} $$\frac{U_{\1 \2}}{U_{\2 \1}} = 
\frac{e^{\mu_\1(\rho_\1)}} {e^{\mu_\2(\rho_2)}}$$ where 
$\mu_\1(\rho_1)$ and $\mu_2(\rho_2)$ are functions of respective 
density. By substituting this ratio in Eq. \ref{LG1} and then 
integrating over densities, one readily obtains the joint 
distribution $P(M_\1,M_\2)$ of particle numbers $M_\1$ and $M_\2$, 
which has exactly the same form as given in Eq. \ref{LG2}. Then, 
minimizing total free energy function, one can identify 
$\mu_\1(\rho_\1)$ and $\mu_\2(\rho_\2)$ as chemical potentials 
which equalize in the final steady state after the systems are 
brought into contact; the equalization of this chemical potential 
was indeed verified through simulations in \cite{Pradhan2_PRE2011}. 
However, the microscopic exchange rates $u_{\alpha \alpha'}$ have 
not been derived from the canonical fluctuation-response relation 
Eq. \ref{FR1} and therefore are not constrained by the balance 
condition Eq. \ref{Balance_intro}. Consequently, as in the 
previous example, these exchange rates lead to the breakdown 
of ensemble equivalence. That is, free energy function 
and chemical potential for systems in grandcanonical ensemble 
are not the same as those for isolated systems in canonical 
ensemble.

\subsection{Lattice gases with nearest neighbour exclusion}

Next we consider previously studied athermal hardcore lattice 
gases, in two dimensions, with nearest neighbour exclusion (NNE) 
\cite{Dickman_PRE2014, Dickman2_PRE2014}. We study the simplest 
case where particles can be exchanged through a single-site or 
point-wise contact ($v=1$) in each system, which can be readily 
generalized to other cases, e.g, when particles are exchanged 
globally ($v=V$) or in higher dimensions. The transition rates for 
particles hopping inside the 
individual systems (irrespective of that they are isolated or in 
contact with each other) can be chosen to be some specific nearest 
neighbour or next-nearest neighbour (or mixture of both) hopping 
rates in the presence of a driving field $D$; details of these 
rates, which can be found in \cite{Dickman_PRE2014, 
Dickman2_PRE2014}, are omitted here as they are not explicitly 
required in the following analysis as long as the systems 
exchange particles very slowly.

Let us keep two such lattice gases, systems $\alpha=\1$ and $\2$, 
in contact with each other \cite{Dickman_PRE2014, Dickman2_PRE2014} 
where particles are exchanged as follows. A site is called open if 
the site as well as all its nearest neighbours are unoccupied. 
Provided the contact site, say in system $\1$, is occupied and the 
contact site in system $\2$ is open, the particle from system $\1$ 
is transferred to system $\2$ with rate $u_0 \rightarrow 0$. The 
joint distribution $P(M_\1, M_\2)$ of masses $M_\1$ and $M_\2$ in 
the individual systems can be straightforwardly 
calculated by substituting $U_{\alpha \alpha'}(M_\alpha, \mDelta) = 
\rho^c_\alpha \rho_{\alpha'}^{\rm c, op}$ (with $\mDelta=1$) in Eq. 
\ref{LG1} where $\rho^c_\alpha$ and $\rho_{\alpha'}^{\rm c, op}$ 
are probabilities that contact site is occupied in system $\alpha$ 
and open in system $\alpha'$, respectively. Note that the 
probabilities 
$\rho^c_\alpha({\bf x_c}, \rho_\alpha)$ and $\rho_{\alpha}^{\rm c, 
op} ({\bf x_c}, \rho_\alpha)$ are, in principle, functions of the 
location ${\bf x_c}$ of the contact site as well as of the global 
density $\rho_\alpha$ in system $\alpha$.

Then the joint distribution has the same form as given in Eq. 
\ref{LG2} where free energy densities can be written as
$f_\1(\rho_\1) = \int_0^{\rho_\1} \mu_\1 d\rho_\1$ and $f_\2 
(\rho_\2) = \int_{M/V_1}^{\rho_\2} \mu_\2 d\rho_\2$ with chemical 
potentials given by \be \mu_\alpha (\rho_\alpha) = \ln \left( 
\frac{\rho^c_\alpha}{\rho_\alpha^{\rm c, op}} \right). \ee
The macrostate is obtained by minimizing total free energy 
function $F = V_\1 f_\1(\rho_\1) + V_\2 f_\2(\rho_\2)$ with the 
constraint $V_\1 \rho_\1 + V_\2 \rho_\2 = {\rm constant}$, leading 
to the existence of an intensive thermodynamic variable, i.e., 
chemical potential, which indeed equalizes upon contact, 
$\mu_\1(\rho_\1) = \mu_\2(\rho_2)$. However, the functional 
form of the chemical potentials do depend on the boundary 
conditions. Because, a 
particular boundary condition can make the density profile 
nonuniform and, consequently, the quantities $\rho^c_\alpha ({\bf 
x_c}, \rho_\alpha)$ and $\rho_{\alpha}^{\rm c, op} ({\bf x_c}, 
\rho_\alpha)$ not only depend on density $\rho_\alpha$ but also on 
the location ${\bf x_c}$ of the contact site.

For example, in the case of periodic boundary condition and 
uniform bulk hopping rates where the system remains homogeneous, 
chemical potential is 
given by \be \mu_\alpha = \ln \left( \frac{\rho_\alpha}{\rho^{\rm 
op}_\alpha} \right), \ee where the density $\rho^c_\alpha ({\bf 
x_c}, \rho_\alpha) = \rho_\alpha$ at the contact site ${\bf x_c}$ 
and the  probability $\rho^{\rm c, op}_\alpha ({\bf 
x_c}, \rho_\alpha) = \rho^{\rm c, op}_\alpha(\rho_\alpha)$ of the 
contact site being open depends only on the bulk density 
$\rho_\alpha$, i.e., both $\rho^c_\alpha$ 
and $\rho^{\rm c, op}_\alpha$ do not depend on the location ${\bf 
x_c}$ of the contact. This is exactly the chemical potential which 
was found in \cite{Dickman_PRE2014}, using the concept of virtual 
exchange, for the pointwise (single-site contact with $v = 1$) 
as well as for global exchanges ($v = V_\alpha = V_{\alpha'}$).

On the other hand, for hard-wall or reflecting boundary condition 
(e.g., periodic boundary in $x$ direction and two hard walls placed 
along $x=1$ and $x=L$), the density profile becomes nonuniform and 
the chemical potential then depends on where the contact site is 
located. For example, if the contact site is located in the bulk,
chemical potential has to be calculated with respect to the 
density and probability of open site in the bulk. That is, even in 
these cases of nonuniform systems, the existence of the above 
mentioned chemical potential would then apparently restore an 
equilibrium-like thermodynamic structure, as formulated in 
\cite{Dickman_PRE2014, Dickman2_PRE2014} where an ITV equalizes 
upon contact and zeroth law is obeyed.

In short, in all the above cases of weakly interacting NNE lattice 
gases with uniform or nonuniform density profiles, there indeed 
exists, in the limit of slow exchange, an ITV which equalizes upon 
contact and zeroth law is also obeyed. However, in each of these 
cases - depending on the boundary conditions and the location of 
contact site, the functional form of free energy and chemical 
potential of the individual systems in the grandcanonical ensembles 
are different. Of course, they are not the same as those defined 
for the individual isolated systems in canonical ensemble.

\section{Summary and Discussion}

In this paper, we demonstrate that {\it weakly interacting} 
nonequilibrium systems, with short-ranged spatial correlations and 
having a common conserved quantity, e.g., mass which is exchanged 
upon contact between two systems, have an equilibrium-like 
thermodynamic structure in steady state, provided the rates of mass 
exchange between two systems satisfy a balance condition as given 
in Eq. \ref{Balance_intro}. The size of the contact regions, 
otherwise arbitrary, should be much larger than correlation 
lengths, therefore making the contact regions effectively 
independent of the rest of the systems. The balance condition, 
reminiscent of equilibrium detailed balance on a coarse-grained 
level, leads to zeroth law of thermodynamics and fluctuation-
response relations analogous to the equilibrium fluctuation-
dissipation theorems. In other words, for mass exchange rates 
satisfying the balance condition, one can construct equivalence 
classes consisting of systems having a nonequilibrium steady state. 
The systems in each class are specified by the value of an
intensive thermodynamic variable, inherently associated with the 
respective isolated systems, which does not change when any two
systems in the class are allowed to exchange mass according to Eq.
\ref{Balance_intro}.

Following are the two most important aspects in the present study. 
Firstly, we constructed a well-defined thermodynamic structure, 
encompassing all (driven or nondriven) steady-state systems 
having {\it nonzero}, though short-ranged, spatial correlations. 
Secondly, we have identified 
the notion of {\it weak interaction} in constructing such a 
thermodynamic structure. Note the distinction between the limit of 
weak interaction and the limit of mere slow mass exchange; the 
former essentially implies vanishing of spatial correlations 
between two systems while in contact (ensuring that there is no 
inhomogeneities at the contact regions) and, moreover, leads to the 
additivity property as formulated in Eq. \ref{additivity1}, 
provided that the balance condition Eq. \ref{Balance_intro} is
satisfied.

In equilibrium, weak interaction directly translates 
into vanishingly small interaction energy between two systems 
in contact, i.e., sum of the internal energies of the individual 
systems equals to total internal energy of the combined system. 
However, in nonequilibrium, the microscopic weights are not 
determined by energy function and therefore even zero interaction 
energy could lead to nonzero spatial correlations between two 
systems while in contact, e.g., when mass exchange rates are finite 
or nonuniform. In principle, the weak-interaction limit can be 
achieved by keeping the bulk transition rates (i.e., the internal 
dynamics in the individual systems) unchanged, irrespective of 
whether the systems are in contact with each other or they are 
isolated. Weak interaction, which usually requires slow exchange of 
masses, is possible even when mass exchange rates are finite, e.g., 
when the balance condition Eq. \ref{Balance_intro} holds.

This thermodynamic construction, which is based on additivity property, may not be valid for the systems having a slow decaying long-ranged spatial correlation, e.g., two-point correlation function decaying as $1/r^d$ (or slower) in $d$ dimensions, which has been observed in a large class of driven systems \cite{Garrido, Grinstein}. In that case, the correlation function is not integrable and therefore the additivity property in Eq. \ref{factorw} presumably breaks down, implying that the fluctuation-response relation in Eq. \ref{FR1} may not exist. Nevertheless, as we demonstrated in this paper, the results will be applicable to a still wide class of driven systems which have short-ranged correlations. Moreover, even in the presence of long-ranged correlations when the strength of the correlations is weak, the additivity property, to a good approximation, could hold. This possibly explains why driven lattice gases, such as KLS models studied in Refs. \cite{HayashiSasa_PRE2003, SasaTasaki_JStatPhys2006, Pradhan_PRL2010, Dickman_PRE2014}, admit an approximate free energy and chemical potential, thus providing a quite good description of various steady state properties - including description of phase transitions \cite{Pradhan2_PRE2011} - albeit only in the limit of weak interaction.

It is important to note that slow exchange of masses does not 
necessarily imply weak interaction. For example, the nonuniformly 
driven athermal lattice gas studied in \cite{Dickman2_PRE2014} is 
one where the system is {\it not} actually weakly interacting, even 
when mass exchange rates are vanishingly small or slow. In a 
realistic scenario, finite interaction may be present between two 
systems while in contact. As an open issue, it remains to be seen 
whether, in the case of finite interaction, there exists an 
intensive thermodynamic variable which would equalize upon contact. 
Also, it would be interesting to explore the validity of additivity  
property in systems having boundary layers or hard walls, as their 
presence could alter the fluctuations in the bulk of a system which 
is otherwise isolated. A related important open question 
\cite{SasaTasaki_JStatPhys2006} is whether the thermodynamic 
structure based on additivity could be used to connect various 
physical observables, such as mechanical pressure on a wall 
\cite{Solon1, Solon2} or statistical forces on a probe \cite{Basu}, 
to an intensive thermodynamic variable such as chemical potential. 
Though addressing the issue in full generality remains a 
formidable challenge, it would be worthwhile to identify a 
particular class of driven systems, if any, where connection 
between ‘mechanics’ and nonequilibrium thermodynamics could be 
established on a firmer ground.

We end the discussion with a concluding remark. The problem 
of constructing a well-defined thermodynamic structure in 
nonequilibrium, even when spatial correlations are short-ranged, is 
more subtle than that in equilibrium as, in nonequilibrium, zeroth 
law alone cannot ensure an equivalence class. Even when zeroth law 
holds, nonequilibrium ensembles (canonical and grandcanonical) may 
not be equivalent as the fluctuation properties of systems in 
grandcanonical ensemble depend on the details of contact dynamics 
as well as the boundary conditions - which gives insights into the 
conceptual difficulties in constructing a nonequilibrium 
thermodynamics, e.g., as attempted in 
\cite{SasaTasaki_JStatPhys2006, Dickman_PRE2014, Dickman2_PRE2014, 
Pradhan_PRL2010, Pradhan1_PRE2011}. In this scenario, our study 
provides a general prescription for dynamically generating 
different equivalent nonequilibrium ensembles and could thus help 
in formulating a well-defined nonequilibrium thermodynamics for 
driven systems in general.

\section{Acknowledgement}

SC acknowledges the financial support from the Council of Scientific and Industrial Research, India [Grant No. 09/575(0099)/2012-EMR-I]. PP and PKM acknowledge the financial support from the Science and Engineering Research Board (Grant No. EMR/2014/000719).

\end{document}